\documentclass[preprint2]{aastex}

\slugcomment{ submitted to ApJ}

\shorttitle{Abell 754}
\shortauthors{Kale \& Dwarakanath}

\begin{document}

\title{Diffuse radio emission in Abell 754}

\author{Ruta Kale \& K. S. Dwarakanath}
\affil{Raman Research Institute, Bangalore-560080}

\email{ruta@rri.res.in, dwaraka@rri.res.in}

\begin{abstract}
We present a low frequency study of the diffuse radio emission in the galaxy cluster A754.
We present new 150 MHz image of the galaxy cluster A754 made with the Giant Metrewave Radio Telescope (GMRT) and discuss the detection of 4 diffuse features. 
We compare the 150 MHz image with the images at 74, 330 and 1363 MHz; one new diffuse feature is detected. 
The flux density upperlimits at 330 and 1363 MHz imply a synchrotron spectral index, $\alpha > 2$, ($S\propto \nu^{-\alpha}$) for the new feature.
The 'west relic' detected at 74 MHz (Kassim et al 2001) is not detected at 150 MHz and is thus consistent with its non-detection at 1363 MHz (Bacchi et al 2003) and 330 MHz(Kassim et al 2001).
Integrated spectra of all the diffuse features are presented. 
The fourth diffuse feature is located along the proposed merger axis (Zabludoff et al 1995) in A754 and 0.7 Mpc away from the peak of X-ray emission; we refer to it as a relic.
We have made use of the framework of adiabatic compression model (Ensslin \& Gopal-Krishna 2001) to obtain spectra. We show that the spectrum of the fourth diffuse feature is consistent with that of a cocoon of a radio galaxy lurking for about $9\times10^{7}$ yr; no shock compression is required. The other three diffuse emission have spectra steeper than 1.5 and could be cocoons lurking for longer time. We discuss other possibilities such as shocks and turbulent reacceleration being responsible for the diffuse emission in A754.
\end{abstract}
\section{Introduction}
Diffuse, synchrotron radio sources 
with no obvious optical counterparts have been detected  
in several clusters of galaxies. They provide evidence for the existence of $\sim$ Mpc 
scale distribution of magnetic fields and relativistic particles in the intra-cluster medium (ICM). These sources are broadly classified into two types- radio halos and radio relics.
Radio halos are unpolarized ($\lesssim5\%$), have a regular morphology and are centrally located in galaxy clusters like the hot gas emitting thermal X-rays; radio relics are elongated, single or double arc-like, located at the peripheries of clusters and show strong polarization ($\sim 20\%$) (Feretti \& Giovannini 1996; Giovannini et al 2002; Ferrari et al 2008). 
These sources have steep spectra ($\alpha > 1$, $S \propto \nu^{-\alpha}$) and are $\sim0.5$ - 1 Mpc in size. Halos and relics have typical surface brightnesses of $\sim $mJy arcmin$^{-2}$ at 1.4 GHz. The origins of relativistic electrons and large scale magnetic fields in the ICM are still unclear. The relativistic electron population in the ICM could either be injected by radio galaxies and stellar feedback or electrons in the ICM accelerated in shocks. The relativistic (Lorentz factor $\gamma \gg 10^{4}$) population of electrons cannot diffuse to distances of $\sim$ Mpc in their short radiative lifetimes of $\sim10^{8}$yr in magnetic fields $\sim0.1 - 1\mu$G. This leads to the requirement of in-situ re-acceleration of electrons in the ICM to produce a Mpc scale radio halo. 
There are two classes of models which are considered - (a) re-acceleration of electrons by turbulence  in the ICM (reviews by Brunetti  2003, 2004; Sarazin 2004; Fujita et al 2003; Brunetti et al 2004; Brunetti \& Lazarian 2007) and (b) the secondary electron models(Dennison 1980; Blasi \& Colafrancesco 1999; Dolag \& Ensslin 2000). In the secondary electron models, the protons in the ICM that are accelerated to relativistic energies in shocks lead to the production of relativistic electrons and positrons by inelastic p-p collisions . The secondary electrons are produced continuously in the cluster by the protons which have lifetimes comparable to the age of the cluster and thus cluster wide emission can be explained without encountering the problem of short radiative lifetimes of relativistic electrons. The gamma ray flux from the ICM predicted by these models has not been detected so far. This is a serious drawback of these models. 
The two models considered for relics are (a) diffusive shock acceleration of thermal electrons by Fermi I process (Ensslin et al 1998; Roettiger et al 1999) or of fossil relativistic electrons (Markevitch et al 2005) and (b) revival of fossil radio cocoons by adiabatic compression due to shocks in the ICM (Ensslin \& Gopal Krishna 2001, hereafter EG01; Ensslin \& Bruggen 2002; Hoeft et al. 2004).
\par Large scale structure formation simulations show that structure formation, accretion and mergers drive shocks in the ICM (Miniati et al. 2000; Ryu et al. 2003; Pfrommer et al. 2006). Sub-structures in the X-ray emission from cluster cores, imaged in high resolution by {\it Chandra} and XMM-Newton, have indicated non-virialized states of galaxy clusters (eg. McNamara et al. 2002; Markevitch et al. 2003). 
The gravitational binding energy released in cluster-cluster mergers is $\sim 10^{63}$ erg and thus if only $\sim1 \%$ of it is utilised in accelerating particles, sources like the radio halos ($\sim 10^{60}-\sim 10^{61}$ erg) could be produced. 
There is increasing evidence that radio halos and relics occur in clusters which have undergone a merger event in the last $\sim 1$ Gyr (Feretti et al. 2006). 
In a few clusters such as A2256 (Briel \& Henry 1994) and A1367 (Donnelly et al. 1998) it has been possible to detect cluster merger shocks as temperature jumps in the X-ray emitting ICM. These clusters are also hosts to radio relics which have been proposed to have formed by acceleration of electrons in the shock structures of the cosmological large-scale matter flows (Ensslin et al 1998).
The shocks or turbulence in the ICM, which are considered responsible for the acceleration of electrons, would leave signatures in the resulting spectrum of the accelerated particles. 
Spectral index is a tracer of the age of the relativistic plasma and hence its distribution over the halo/relic can reveal sites of ongoing reacceleration.
Relics showing a trend of gradual flattening of spectral index from the edge near the cluster core to the edge farther from the cluster core could indicate relation to outgoing merger shocks. 
In the case of the double relics in A1240 (Bonafede et al. 2009) and the relic in A521 (Giacintucci et al. 2008)  this possibility based on morphologies of the relics, the spectral index maps and the polarization has been considered.
Patches of steep/flat spectral index could indicate complex process of re-acceleration 
such as turbulent reacceleration. 
 Radial steepening of spectral index of halos could also be related to the cluster scale magnetic field distribution (Feretti et al 2004). 
The correspondence between X-ray sub-structure or temperature map and the spectral index maps can reveal properties of the relativistic electron population. 
In order to carry out such studies of halos and relics, multi-frequency observations are necessary. 
\par 
Radio halos and relics have been studied mainly at 1.4 GHz (Giovannini et al 1999; Govoni et al 2001; Bacchi et al 2003(hereafter BA03); Clarke et al 2006). 
Observations at frequencies other than $1 $GHz are required to study the spectra of halos and relics and to compare with model predictions.
 Recently a low frequency halo has been detected in the cluster A521 by Brunetti et al. (2008) in the maps at 235, 330 and 610 MHz. This halo was missed at higher frequencies ($\gtrsim 1$ GHz) due to cutoff in the spectrum. This detection emphasises the importance of low frequency observations and provides an example of a class of halos which will be detected only at low frequencies. The extended nature and steep spectra make halos and relics ideal targets for low frequency ($<1$ GHz) observations.
In recent years, the Giant Metrewave Radio Telescope (GMRT) has provided imaging opportunities at frequencies of 610, 330, 235 and 150 MHz with resolutions of $\sim 5''$, $9''$, $13''$ and $20''$ respectively. 
A few attempts have been made to image halos and relics with the GMRT.
The radio halo in galaxy cluster A3562 has been successfully imaged at 235 MHz by Giacintucci et al (2005). Brunetti et al. (2008) have discovered the low frequency halo in A521 with the GMRT. A search for radio halos and relics in galaxy clusters, which resulted in the detection of 10 new radio halos and 3 relics, has been carried out at 610 MHz with the GMRT (Venturi et al 2008). At low frequencies, the primary beams are larger compared to those at higher frequencies and thus a larger field can be imaged in a single pointing. Considering the fact that relics occur at peripheries of clusters, it is essential that a large field around the cluster centre be imaged.
For example, in A2255, which is well known for the presence of a halo and relic, new relics at a distance of $\sim 2$ Mpc ($\sim 25'$) from the cluster centre were discovered in 350 MHz observations (Pizzo et al 2008). 
\par With the hybrid configuration of the GMRT, imaging of structures of angular sizes $\sim 1^{\circ}$ with a resolution of $\sim 20''$ at 150 MHz and with a sensitivity of $\sim$1.5-3 mJybeam$^{-1}$ is possible. 
We selected halos and relics for low frequency observations with the GMRT. 
Haloes/relics which  had extents of $10'-30'$ at 1.4 GHz were selected so that the coverage at short baselines available at the GMRT is sufficient to image the full extents of these sources.
Further, sources with expected $1\sigma$ surface brightness at 150 MHz of $>1.5$mJybeam$^{-1}$ were selected.  
With our previous experience of the dynamic range limitation of $\sim$ 1500 in the GMRT images, we avoided those clusters which had radio sources $\gtrsim3 $ Jy within the HPBW of primary beam ($\sim 3^{\circ}$) at 150 MHz.
  \par 
As a first phase of imaging galaxy clusters at low frequencies, three galaxy clusters  A2255, A2256 and A754 were observed  
at 150 MHz with the GMRT. Galaxy cluster A2255 is host to a radio halo and a relic (Feretti et al. 1997) and two relics about 2 Mpc from the cluster centre (Pizzo et al 2008). The radio halo of A2255 is polarized ($\sim 20\%$, Govoni et al 2005). In A2256 a radio halo and a relic have been detected (Kim 1999; Clarke et al 2006; Brentjens 2008). In A754 two components of diffuse radio emission (Kassim et al. 2001, hereafter KA01; BA03) have been detected but the nature of these are still unclear. 
The X-ray images of these clusters show sub-structures which indicate on-going mergers
(Sun et al 2002; Markevitch et al 2003; Sakelliou et al 2006 ). These are of particular interest to investigate the connection between the mergers and the radio halos/relics.
 Apart from being hosts to interesting extended diffuse radio sources, these clusters are complex and dynamically active.
 In this paper we present our results for the galaxy cluster A754. 
\subsection{Abell 754}
A754 is a rich galaxy cluster at a redshift of 0.0542 \cite{strub99}, with an X-ray luminosity of $6.4\times10^{44}$ $h_{75}^{-2}$ erg s$^{-1}$ measured in the ASCA energy band of 0.5 - 10 keV (Henriksen \& Markevitch 1996) and a temperature of $9.0\pm 0.3$keV (Wu et al. 1998; Govoni et al 2004). 
Non-virialized nature of A754 has been inferred from the morphologies of X-ray emissions in the images made with ROSAT (Henry \& Briel 1995), {\it Chandra} (Markevitch et al 2003) and XMM-Newton (Henry 2004).
The projected  distribution of optical galaxies has been reported to be bimodal and oriented along the east-west axis; the mean redshifts of the two galaxy concentrations being identical within errors (Fabricant et al 1986; Zabludoff \& Zaritsky 1995, hereafter ZZ95). The peak of X-ray emission in A754 is not spatially coincident with either of the two major galaxy concentrations (ZZ95).
A merger of two sub-clusters approximately in the plane of the sky along the east-west axis has been suggested (ZZ95). Henriksen et al. (1996) using {\it ASCA} data proposed a non-head-on collision of subclusters in A754.
Roettiger et al (1998), using 3D MHD/N-body simulations, have proposed explanations for most of the observed X-ray properties of A754 by considering an off-axis merger between 2 smaller clusters of mass ratio less than 2.5:1. 
Markevitch et al. (2003), based on the temperature map obtained using the {\it Chandra} data on A754, have suggested a complex geometry of merger involving more than 2 subclusters or a cool gas cloud sloshing independently from its former host subcluster. Fusco-Femiano et al (2003) have reported a 3.2$\sigma$ detection of excess X-ray emission at energies above 45 keV with the {\it BeppoSAX} which is possibly due to the inverse-Compton emission by the relativistic plasma in the ICM of A754.
Absence of cooling flow (White et al 1997; Peres et al 1998) in A754 is another property indicating the on-going dynamical activity in the cluster.
\par The diffuse radio emission in A754 has been under investigation for more than 25 years. Diffuse radio emission in A754 at 2.7 GHz (Wielebinski et al 1977) and at 408 MHz (Mills, Hunstead \& Skellern 1978) have been reported. The diffuse emission could not be distinguished from the blend of discrete sources due to poor resolutions of $3'-4'$ in those images.
The presence of diffuse halo source in A754 at 610 MHz was also suggested by Harris et al. (1980) based on the deficit in the flux recovered from the emission of discrete sources compared to the total flux reported in earlier low frequency observations. Attempts of investigating the diffuse radio emission at 330 MHz in a 40 minute observation with the VLA in B and C arrays (Giovannini, Tordi \& Feretti 1999; Giovannini \& Feretti 2000) failed to detect the diffuse emission due to lack of uv-coverage. The presence of diffuse emission in A754 has been confirmed by KA01 in the images made from 3 hr observations at each of 330 and 74 MHz with the VLA-C array. 
In the deepest image of A754 at 1.3 GHz with the VLA-D array, BA03 have confirmed the presence of two components in the diffuse radio emission.\\
\par In this paper we use $\Omega_{m}=0.27$, $\Omega_{\Lambda}=0.73$ and a value of 73 km s$^{-1}$Mpc$^{-1}$ for the Hubble constant, $H_o$. This implies a distance scale of 61.9 kpc arcmin$^{-1}$ at the redshift of A754.
 \section{Observations and Data Analysis}
GMRT is an aperture synthesis radio telescope consisting of 30 dish antennas, each 45 m in diameter. 
The array has 14 dishes placed in an area of 1 km$^{2}$, forming a central compact array, providing the uv- coverage at short baselines($\sim 100$m).
 Remaining 16 antennas are spread along the 3 arms of a Y-shape to a maximum distance of $\sim25$ km. 
GMRT observations of the A754 region at 150 MHz for a total of $\sim5$hr were conducted on 17 Sep. 2007. 
High resolution (HIRES) mode of the GMRT correlator provided frequency resolution of 31.25 kHz over a bandwidth of 8 MHz.
This mode was used to minimize the effects of bandwidth smearing and for better excision of radio frequency interference(RFI). 
Standard sources 3C48 and 3C147 were used for the flux and the bandpass calibration. The source 0837-198 was used as the gain and phase calibrator.
\par Data at 150 MHz with the GMRT were adversely affected by RFI. 
An RFI affecting about 10 consecutive frequency channels and drifting over 2 MHz in frequency during the total 5 hr of observation was found along with other RFIs which affected certain channels for limited amount of time.
The AIPS (Astronomical Image Processing System) task SPFLG was used to remove channels affected by RFI. A careful examination of frequency channels in all baselines was carried out to prevent loss of any good data while removing the RFI affected data. 
About 35\% data were lost due to RFI. 
Polyhedron imaging with 66 facets was carried out to account for the non-coplanarity of the incoming wavefront within the large primary beam of $\sim 3^{\circ}$. Hydra A, a strong radio source, $\sim 3.5^{\circ}$ from the phase centre was imaged to remove the effect of its sidelobes from the field of interest.
The FWHM of the primary beam of the GMRT at 150 MHz is $\sim3^{\circ}$ which corresponds to $\sim10$ km of ionosphere at a height of $200$ km from the surface of the earth. 
The passage through the ionosphere 
of the incoming wavefront leads to an excess ionospheric phase. This excess phase, if constant over the entire field of view, can be solved for using self-calibration. This is the assumption of isoplanicity.
Isoplanicity could be a poor assumption for the GMRT at frequencies lower than 325 MHz due to long baselines and large field of view (Rao 2003). 
Spatial and temporal variations in the electron density in the ionosphere result in variations in  excess phase within the primary beam. This leads to variations in the positions of sources with time.
Furthermore, at any given time the different positions in the primary beam can experience different position shifts. This is non-isoplanicity. One self calibration solution may not work for the entire primary beam under such conditions. 
We made images of 10 min data selected at the 
beginning and at the end of the observation to check if there was any effect of non-isoplanicity.
Positions of the sources in the entire primary beam in both the 10 min images were consistent with each other and also with 
those in the NRAO VLA Sky Survey(NVSS) catalogue within $1\sigma$ error of $3''$. 
Self calibration worked equally well for all the sources in the entire primary beam. No effects of non - isoplanicity were seen.\\
\par Data at 330 MHz from the GMRT archive were also edited and calibrated using AIPS. At 330 MHz  $\sim 20\%$ data were lost due to RFI. Polyhedron imaging was carried out with 32 facets. Standard steps of self calibration were performed. Image with a resolution of $10''\times10''$ was produced. We used the VLA D-array data from the archive at 1363 MHz to make an image with a resolution of $50''\times50''$. These images using the archival data were used to estimate the contributions of unresolved sources to the diffuse emission.
Images at 74 and 330 MHz(KA01) with resolutions of $316''\times233''$ and $75''\times56''$, respectively, were provided by N. Kassim and the image with a resolution of $70''\times70''$ at 1363 MHz(BA03) was provided by G. Giovannini. These images were used to study the diffuse emission.
Details of the data are given in Table 1.
\section{Results}
GMRT image with the highest possible resolution of $21''$ was made using the 150 MHz data.
While this high resolution image showed the detailed morphologies of several cluster radio galaxies, images with lower  resolutions were produced at 150 MHz to detect the diffuse halo emission.
 For comparison with the 1363 MHz image, we convolved the image at 150 MHz with a beam of $70''\times70''$; rms of 5.5 mJybeam$^{-1}$ was achieved.
Images at 150 and 1363 MHz of the A754 region where the diffuse emission is detected are shown in Fig. 1a and 1b respectively. In Fig. 2a and 2b, images at 330 and 74 MHz(Kassim et al. 2001), respectively, are shown. 
In each of these images the locations of diffuse emission detected at 150 MHz are marked by rectangles.
\par We use labels for discrete sources in the A754 field as used in Fig. 3 of BA03. For estimating the flux densities of discrete sources we have made use of images with resolutions of $50''$ and $10''$ at 1363 and 330 MHz respectively.
The source S6 in Fig. 1b is not detected at 150 MHz. It has flux densities of 33.0 and 7.0 mJy at 330 and 1363 MHz respectively which imply a spectral index of 1.0. The non-detection of S6 above the $3\sigma$ level of 16.5 mJy at 150 MHz indicates that the spectrum has turned over. The non-detection of S5 (2.5 mJy at 1363 MHz) at 330 and 150 MHz is consistent with a spectral index of 0.8. Flux density of S1 at 1363 MHz is 3.6 mJy and the non-detection at 330 and 150 MHz implies a spectrum flatter than 0.8. 
\par The details of the flux densities of each blob of diffuse emission marked by rectangle in Fig. 1a are given in Table 2. 
The 'west relic' in the 74 MHz image of KA01 at RA=09h07m25s DEC=-09d36m52s is not detected in any of the 150, 330 and 1363 MHz images. If the 'west relic' is real then the non-detection of it at 150 MHz implies a spectral index between 74 and 150 MHz steeper than 1.8.
\par The integrated spectra of each of the blobs 1, 2, 3 and 4 with 3$\sigma$ error bars are plotted in Fig. 3. The flux densities of the blobs at each of the frequencies were estimated using rectangles of the same dimension as shown in Fig. 1a. The $1\sigma$ error on the total flux density at 150 and 330 MHz is $20\%$ and at 1363 MHz is $10\%$. In the case where the emission was not detected, the total flux density recovered in the corresponding box is reported as an estimate of upper limit.
These upperlimits are plotted for blobs 1, 2 and 3 at 330 MHz and for blob 1 at 1.4 GHz.
In the case of blob 3, the discrete source S6 contributed 4.8 mJy to the box shown. This value has been subtracted from the total flux density in box 3 to get the estimate of the flux density of the blob 3.
\par The locations of the diffuse radio emission are marked by boxes on the X-ray image (XMM-Newton, 0.8-2 keV, Henry et al. (2004)) of A754 (Fig. 4). The locations of optical subclusters (ZZ95) are marked by star symbols. The proposed direction of merger of 2 sub-clusters of galaxies in A754 is along the line joining the two optical clumps (ZZ95).
The peak of the X-ray emission is elongated perpendicular to the proposed direction of merger and 
so is blob 4 emission but $\sim 700$ kpc from the X-ray peak. There is no clear indication of a shock in the X-ray images of the region where the relic emission is located. The blob 4 is cospatial with the western clump of optical galaxies.
Based on the comparison of temperature map and radio image of A754, Govoni et al (2004) show a clear anti-correlation between the location of the diffuse radio sources and the regions of hottest X-ray emitting gas in A754.
\par The blob 4 in Fig. 1a is consistent with the ``halo`` marked in the Fig.1 of Fusco-Femiano et al (2003). 
However, we shall refer to the blob 4 as a 'relic' due to 
its elongated morphology and the location away from the peak of the X-ray emission.
A spectral index of $1.4\pm 0.2$ was estimated using a linear fit to the spectrum.
The extent of the relic projected in the plane of the sky is $\sim 350\times400$ kpc$^{2}$. The volume of the relic, estimated assuming a cylindrical geometry with the smaller side of the box as the diameter, is $\sim 0.04$ Mpc$^{3}$. 
The equipartition magnetic field of $0.7\mu$G was computed using standard formulae (Pacholczyk 1970), integrating the radio emission between 10 MHz and 10 GHz, with a radio spectral index $\alpha= 1.4$, and assuming equal energy density in protons and electrons, and a volume filling factor of 1.
\section{Adiabatic compression model}
We have tried to understand the diffuse emission in A754 in the context of the adiabatic compression model (EG01).
Radio galaxies deposit relativistic plasma to the ICM through jets. After the jets turn off, apart from the $PdV$ work done while expanding in the ICM, the relativistic plasma in the radio lobes loses energy 
by synchrotron and inverse Compton cooling on timescales of $\sim10^{8}$yr. This plasma, if compressed adiabatically by shock waves in the ICM, can emit detectable radio emission. This idea of reviving fossil radio lobes by adiabatic compression is discussed in EG01 in detail. The salient features of this model are described here.
\par The rate of change of momentum, ($p$) of relativistic electron is proportional to the magnetic energy density (synchrotron losses), the energy density of the cosmic microwave background radiation (inverse Compton losses) and the adiabatic gain or loss due to the change in the volume ($V$) of the radio plasma.
A ratio of the initial volume of the radio cocoon ($V_{0}$) and the volume at time $t$ ($V(t)$), called the compression ratio is defined as $C(t)=V_{0}/V(t)$. The change in the volume is approximated to be a power law in time, $V(t)=V_{0}(t/t_{0})^{b}$. 
Adiabatic expansion of a spherical volume (radius $r_o$ to $r$) of ultra-relativistic particles of energy $E_o$ leads to the scaling of energy as $E=E_o(r_o/r)=E_o(V/V_o)^{-1/3}$ (eqn. 11.30, Longair 1981) and thus of energy density as $E/V = (E_o/V_o)(V/V_o)^{-4/3}$.
Assuming isotropic adiabatic expansion of magnetized plasma, the magnetic field energy density ($u_{B}$) scales as $u_{B}=u_{B,0}(V(t)/V_{0})^{-4/3}$.
The evolution of the radio plasma is divided into five discrete phases - (0) injection, (1) expansion, (2) lurking, (3) flashing and (4) fading. {\it Phase 0} (Injection): The radio galaxy is active. A large expanding volume is filled with the radio plasma by the jets. The typical timescale over which this phase lasts is $\sim0.015$ Gyr (EG01; Alexander \& Leahy 1987). This expansion is supersonic with respect to the external medium with an index $b=9/5$, with the assumption that there is no gas density gradient in the vicinity of the radio galaxy (EG01; Kaiser \& Alexander 1997).
{\it Phase 1} (Expansion): The jets of the radio galaxy are off; the radio cocoon expands in the surrounding medium due to higher internal pressure compared to the surrounding ICM. The expansion is considered Sedov-like ($b=6/5$) throughout this phase for simplicity (EG01). {\it Phase 2} (Lurking): The radio cocoon reaches pressure equilibrium with the ICM and thus the volume of the plasma remains constant.
{\it Phase 3} (Flashing): If a shock compresses a lurking radio cocoon, there is enhanced radio emission due to adiabatic compression of the magnetic field. {\it Phase 4} (Fading): The compressed cocoon continues to lose energy by synchrotron and inverse Compton losses and fades.
Each phase is characterised by its duration ($\Delta t$), the timescale of expansion ($\tau $) and an index ($b$) that determines the rate of expansion/compression of the radio cocoon. The compression ratio during a phase is related to these parameters by the relation, $C(t)=(1+\Delta t/\tau)^{-b}$.
The synchrotron emission in the {\it i$^{th}$} phase at a given frequency $\nu$ is given by, $L_{\nu i}\propto B_{i}V_{i}\int_{p_{max}}^{p_{min}}dp f_{i}(p)F(\nu/\nu_{i}(p))$ where $\nu_{i}$ is the characteristic frequency, given by, $\nu_{i}(p)=3e B_{i}p^{2}/(4\pi m_{e}c)$ and $F(\nu/\nu_{i}(p))$ is the dimensionless spectral emissivity of a mono-energetic electron in an isotropically oriented magnetic field (Ensslin et al 1999) and $f_{i}(p)$ is the electron spectrum in the {\it i$^{th}$} phase resulting from an initial power-law distribution. 
\par Further, the five phases are discussed under three scenarios. Scenario A: Cocoon at cluster centre- The external pressure provided by the ICM is large and thus the pressure inside the radio cocoon is considered to be only twice the external pressure. The lurking phase lasts only for 0.1 Gyr; the cocoon will not be able to emit detectable radio emission after compression if it is older. Scenario B: Cocoon at cluster periphery- The freshly injected plasma is over pressured compared to the surrounding medium by a factor $\sim 100$. Pressure equilibrium with the surrounding medium is achieved faster. The plasma can be revived by shock compression even after 1 Gyr.
Scenario C: Cocoon near an active radio galaxy(smoking gun)- A radio galaxy which is possibly the source of the relativistic electrons is visible near the relic. This case is illustrated by the example of the relic in Coma by EG01
(See EG01 for details). 
In A754 the relic (blob 4) has been detected at all the 3 frequencies. The model has been used to obtain a spectrum which fits the observed spectrum of the relic best. Scenario A was considered best suited for A754 since the relic is located in a region where the X-ray emission has been detected (Henry et al 2004). The steps carried out to arrive at the model spectra from the adiabatic compression model for the relic are described below.
\par The radio plasma was assumed to be in $i^{th}$ phase and the estimated magnetic field ($B$) and volume ($V$) of the relic were used. Using compression ratio ($C$) for the $i^{th}$ phase, back calculation was carried out to find the $B$ and $V$ for $(i-1)^{th}$ phase successively till Phase 0. The spectrum was obtained for each of the phases. It was checked whether the model spectrum in the $i^{th}$ phase, in which the relic was assumed to be in, was a good fit to the integrated spectrum of the relic. 
If the spectrum was a bad fit to the observed spectrum, the plasma was assumed to be in another phase and the same steps were carried out. 
In a case where the relic was assumed to be in Phase 1, the corresponding model spectrum was flatter than could fit the observed data. In another case where the relic was assumed to be in Phase 2, the spectrum was steeper.
A phase intermediate to 1 and 2 was tried and the best fit to the observed spectrum was obtained. This intermediate phase is referred as ``Phase $1.5$''.
 The model spectra are shown in Fig. 5. These spectra were constructed using the parameters listed in Table 3. The observed values of the flux density of blob 4 at 150, 330 and 1363 MHz with the $3\sigma $ error bars are plotted along with the model spectra. The initial energy distribution of the relativistic electrons was assumed to be a power law with a spectral index of 2.6 ($N=N_0 E^{-2.6}$).
The flux density at 150 MHz was used to normalize the model spectrum. 
\par A good fit was obtained by considering the relic to be in flashing phase (''Phase 3'') too. 
The volume in the injection phase (Phase 0) required in this case to produce the observed size of the relic after compression by shock is $\sim 0.15$ Mpc$^{3}$. This implies that a volume of $\sim 0.15$Mpc$^{3}$ must have been filled with radio plasma by cluster radio galaxies forming a gigantic cocoon. Considering the typical size of a radio galaxy to be 100 kpc (middle value in Fig. 9 of Blundell et al 1999) and assuming it to fill a sphere of that radius with relativistic plasma, at least $\sim35$ such sources are required to be active within a period of $\sim1$Gyr, to fill a giant cocoon of volume 0.15 Mpc$^{3}$. Such a situation is improbable in any typical galaxy cluster.
\section{Discussion}
\par  
In imaging extended emission at radio frequencies crucial roles are played by the uv-coverage, the sensitivity and the field of view. These vary with the observing frequency and the instrument used.
Further, the spectra of halos and relics are not the same over the extent of the source; there can be isolated regions of steep and flat spectral indices. Due to this interplay between the different aspects of the observational limitations and the intrinsic properties of the diffuse emission, multi-frequency observations of halos and relics are necessary to obtain maximum information and to make further  inferences. 
In the case of A754, blob 1 has been detected only at 150 MHz whereas blob 4 is detected at all of 150, 330 and 1363 MHz, but better at 1363 MHz.
 The spectra of blobs 1, 2 and 3 are steeper than that of blob 4. The framework of adiabatic compression model (EG01) has been used to explain the spectrum of the blob 4 in A754.
 It was found that, the relic (blob 4) can be considered as a cocoon of a radio galaxy lurking for $\sim 9 \times 10^{7}$ yr in the ICM. No shock compression was required to produce the spectrum of the relic. 
The consideration of blob 4 as a compressed radio cocoon (Flashing phase) led to an unrealistic situation of the cluster having a gigantic ($\sim 0.15$ Mpc$^{3}$) radio cocoon in the past, which appears unlikely.
 Thus, merger shocks need not be invoked to explain the observed properties of the relic. 
The flux densities at 150, 330 and 1363 MHz imply spectral indices steeper than 1.5 for blobs 1, 2 and 3. Based on the result for blob 4 from the adiabatic compression model, these blobs of diffuse emission are likely to be radio cocoons lurking in the ICM for more than $10^{8}$yr. 
\par In a few clusters that have relics, merger shocks have been considered to be responsible for accelerating electrons (eg. A3376, Bagchi et al. 2006). According to the numerical model of Roettiger et al (1998) in A754, a large subcluster has crossed the main cluster from east to west. 
The 4 diffuse blobs in A754 lie along the east-west axis which is approximately the proposed axis of merger (ZZ95). The proposed location of the eastern shock wave is coincident with the X-ray peak and is close to blobs 1 and 2. The steep spectral indices of blobs 1 and 2 could be explained in the Fermi I acceleration scenario by a low Mach number shock (Mach number $< 2$ for a spectral index steeper than 2.0; Blandford \& Eichler 1987). 
The western shock has been proposed to be at a location $\approx 1.8 h^{-1}_{75}$ Mpc from the X-ray peak (Roettiger et al. 1998; KA01). The blob 4 which is in the western region is only 0.7 Mpc from the X-ray peak and is not coincident with the proposed location of the shock. According to the numerical model the shocks crossed the cluster more than 0.3 Gyr ago which is longer than the radiative lifetime of the electrons observed at 150 MHz. 
In A754, no other relics have been detected in the primary beam at 150 MHz. 
Sensitive polarization study of the diffuse emission has not been carried out to find other signatures of shocks such as alignment of magnetic field in the plane of the shock. 
 Upperlimits on polarized flux density of $9\%$ and $15\%$ have been obtained by BA03 on the western and eastern diffuse emission respectively in A754. 
\par Mergers can create turbulence in the ICM which could then accelerate particles. 
It has been proposed that Fermi acceleration powered by turbulence in the ICM can accelerate particles to relativistic energies (Brunetti et al 2008 and references therein) if the turbulence lasts for more than $10^{8}$yr. 
Particles in the ICM, through resonant scattering off the turbulent waves, can be stochastically accelerated to relativistic energies (Melrose 1980).  
The relation between mergers and turbulent acceleration of particles is complex  and the details are not clear.
The steep spectrum with a cutoff at high frequencies is a signature of turbulent reacceleration (Kuo et al 2003; Brunetti et al 2004; Cassano et al 2005). 
A halo has been detected in A521 at low frequencies such as 235, 330 and 610 MHz (Brunetti et al. 2008) having spectral index of $\sim 2$. 
A high frequency spectral cutoff is suspected by the authors for this halo making it undetectable at 1.4 GHz with the reported sensitivity of their observation.
Turbulent reacceleration model has been invoked to explain the low frequency halo. It is likely that many such diffuse features having cutoff at high frequencies could be lurking in clusters but have not been detected since no low frequency observations have been carried out.
The non-detection of blob 1 at 330 and 1363 MHz indicates cutoff in its spectrum. This implies the possibility of turbulent reacceleration.
The case of blob 4, not requiring shock to explain the radio emission, is also consistent with the picture of turbulent reacceleration. 
\par Another possibility is that the shocks accelerated the protons in the ICM which produced secondary electrons and positrons by hadronic collisions. These secondary particles then produced the observed radio-halo emission. In this case the radio emission can last long after the passage of shock. This scenario will be tested with the detection of gamma ray flux expected from the decay of neutral pions after the p-p collisions (Ensslin et al. 1997; Colafrancesco \& Blasi 1998; Blasi 1998; Dolag \& Ensslin 2000).
\section{Conclusions}
We have presented a GMRT 150 MHz image of the A754 region. We detected 4 blobs of diffuse emission at 150 MHz with a resolution of $70''$. We made high resolution images using the archival data at 330 (GMRT) and 1363 MHz (VLA) to estimate flux densities of discrete sources and used the published images of A754 at 74, 330 and 1363 MHz(VLA) to compare the diffuse features. The 'blob 1' is detected only at 150 MHz. The 'blob 4', is detected at all the three frequencies. The blob 4, lies along the axis of merger in A754 and on the far side of the X-ray peak; thus we refer to it as a relic. Within the framework of the adiabatic compression model (EG01), we found that the blob 4 can be explained as an ageing cocoon of an old radio galaxy in the cluster. The spectra of blobs 1, 2 and 3 of diffuse emission are steeper and thus can be considered as radio cocoons older than the relic. Not requiring shock to explain the radio emission of blob 4, is also consistent with the picture of turbulent reacceleration.
\acknowledgements
We thank Biman Nath and Dipankar Bhattacharya for critical comments on the manuscript. We thank G. Giovannini (1363 MHz) and N. Kassim (74 and 330 MHz) for providing the images of A754.
Giant Metrewave Radio Telescope is run by the National Centre for Radio Astrophysics of the Tata Institute of Fundamental Research. The National Radio Astronomy Observatory is a facility of the National Science Foundation operated under cooperative agreement by Associated Universities, Inc. 

\begin{figure}
\epsscale{1}
\plotone{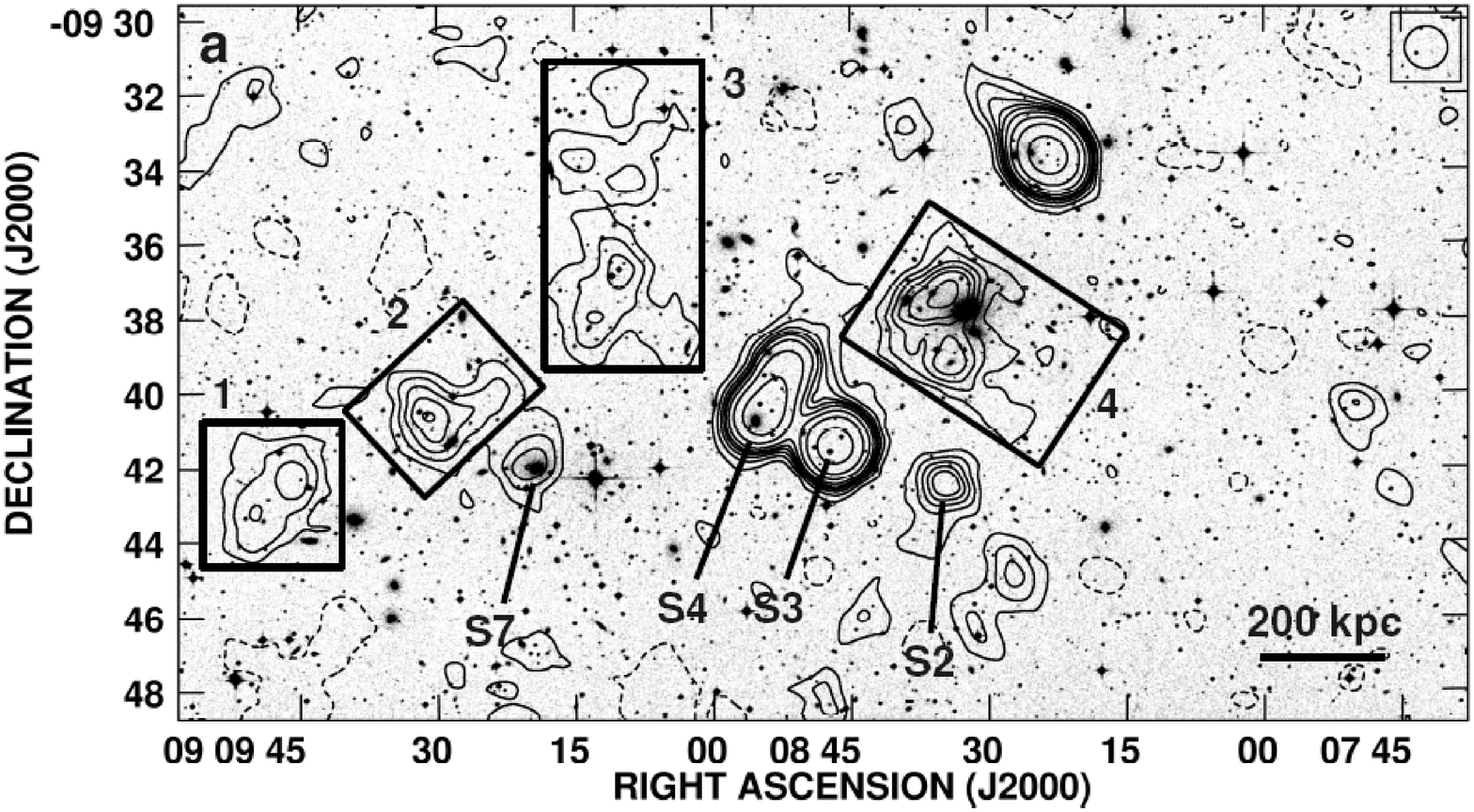}
\plotone{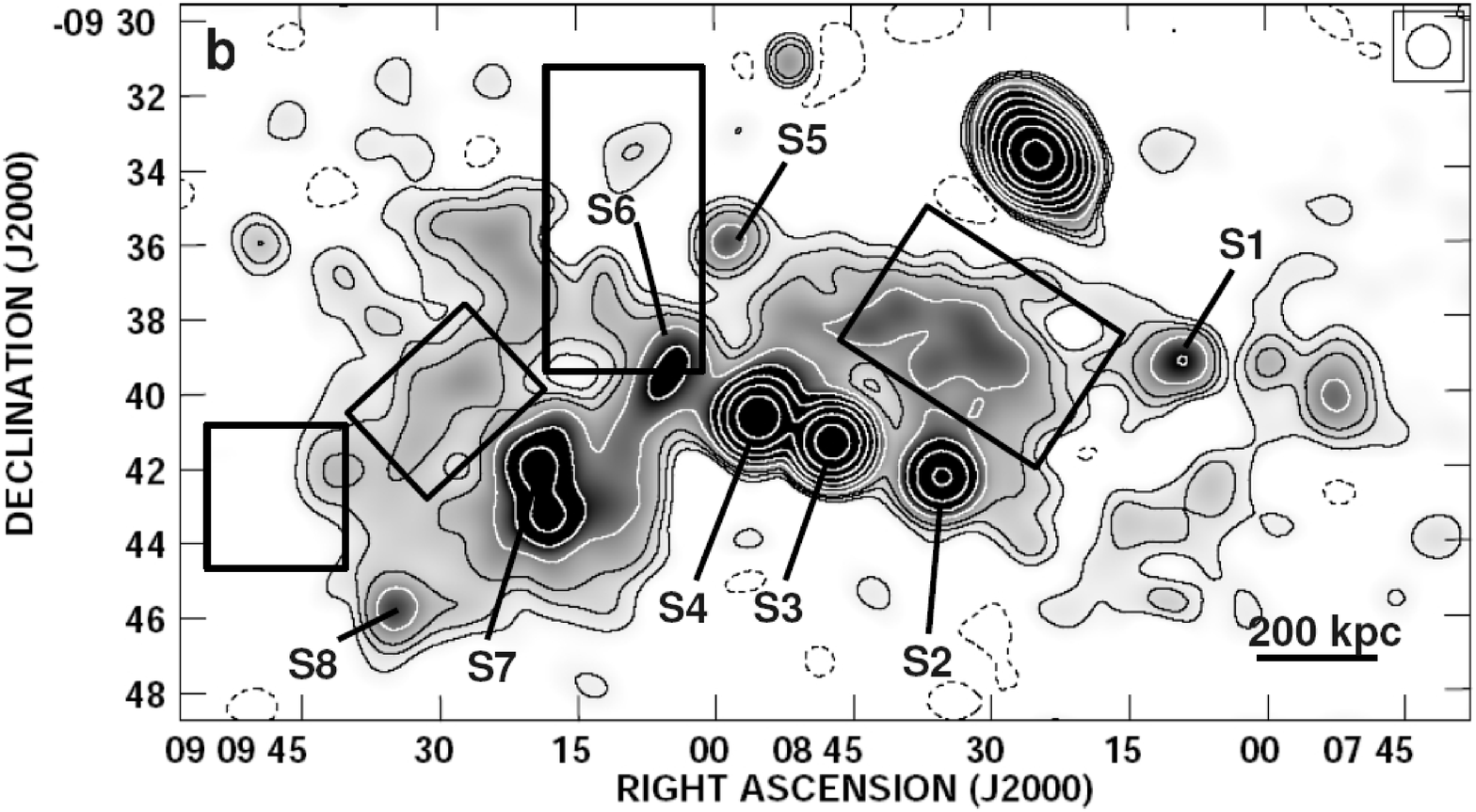}
\caption{A754: (a) Shown in grey-scale is the POSS2 R-band image with the contours at 150 MHz (GMRT) overlaid. 
(b) Image at 1363 MHz is shown in greyscale and contours(courtesy G. Giovannini). The blobs of diffuse emission detected at 150 MHz are marked by rectangles and the discrete sources are indicated by labels in each of (a) and (b). Contour levels: (a) 150 MHz: -16.5, 16.5, 33, 49.5, 66, 82.5, 110, 165, 220, 330 mJybeam$^{-1}$; (b) 1363 MHz (VLA):-0.39,0.39, 0.78, 1.17, 1.56, 1.99, 2.8, 3.9, 7.8, 13 mJybeam$^{-1}$. The convolving beam shown at the top right corner in the images is $70''\times70''$. }{\label{fig 1}}
\end{figure}
\begin{figure}
\epsscale{1}
\plotone{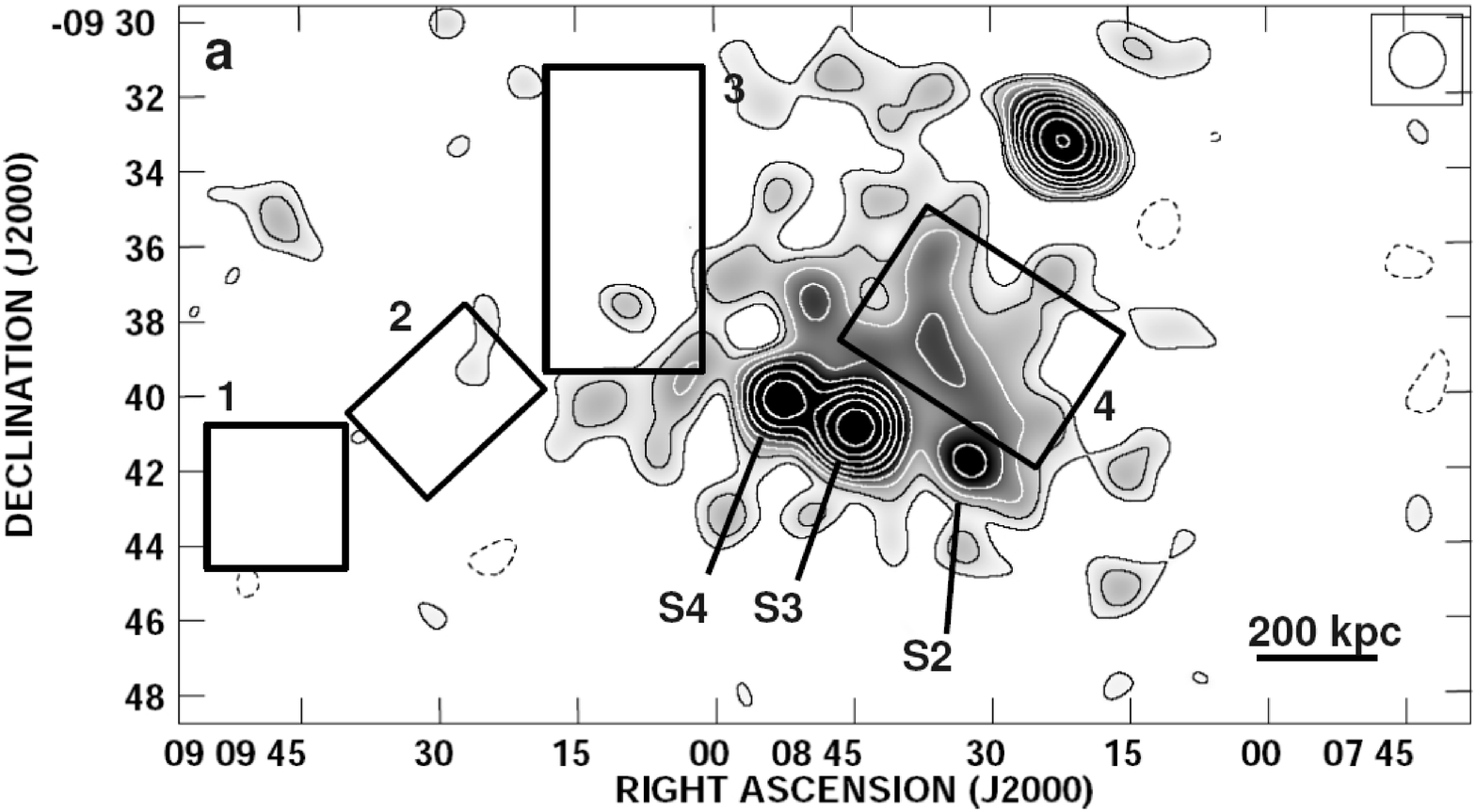}
\plotone{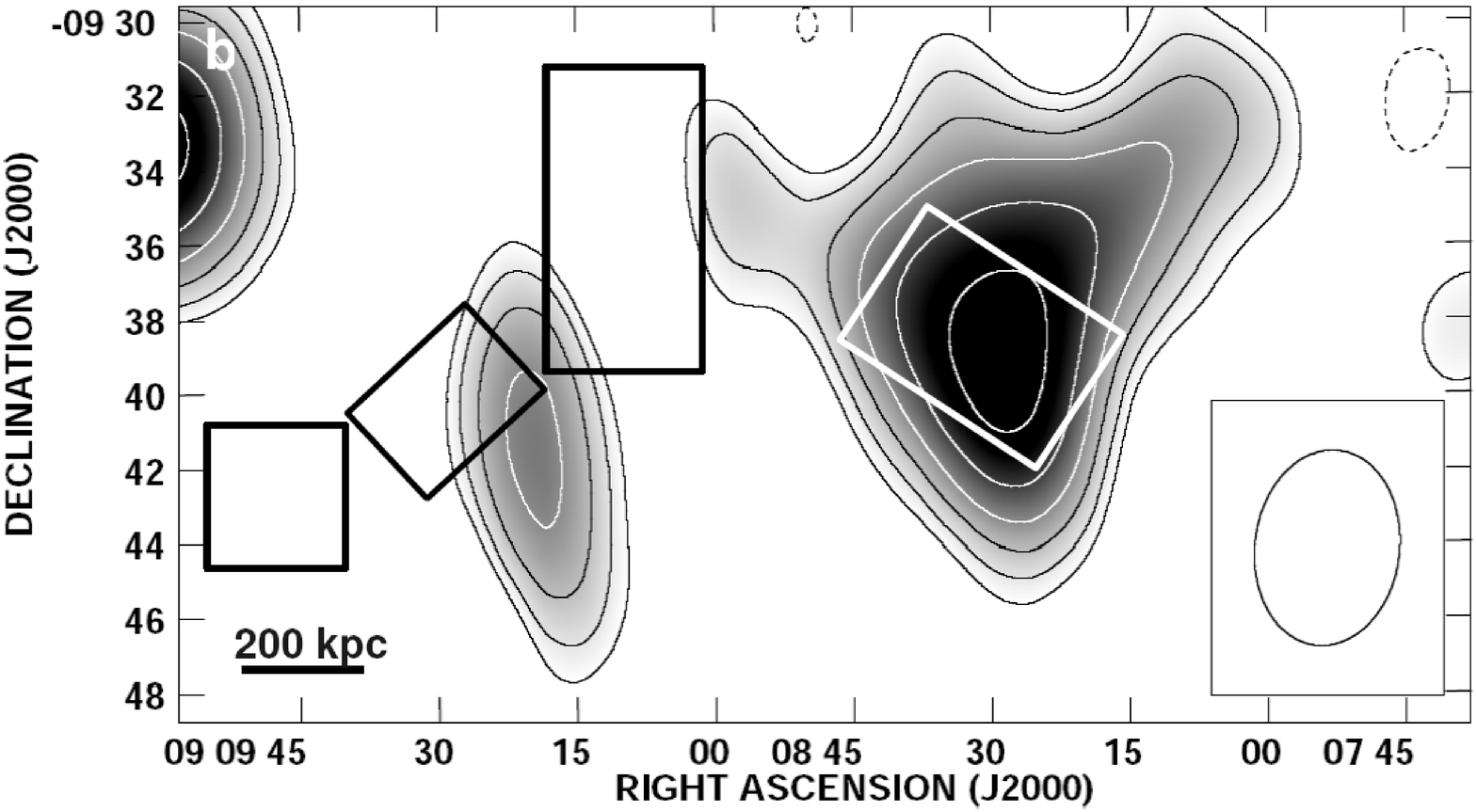}
\caption{In (a) and (b) are 330 and 74 MHz images of A754, respectively, in greyscale and in contours(courtesy Kassim N.). Contour levels are: (a) 16 mJybeam$^{-1}$ $\times$ (-1, 1, 1.4, 2, 2.8, 4, 5.7, 8, 11.3, 16, 22.6, 32) and (b) 0.4 Jybeam$^{-1}$ $\times$ (-1, 1, 1.4, 2, 2.8, 4, 5.7, 8, 11.3). Beam sizes are $90''\times90''$ and $316''\times233''$(P. A.= -7$^{\circ}$) in (a) and (b) respectively.}{\label{fig 2}}
\end{figure}
\begin{figure}
\epsscale{1}
\plotone{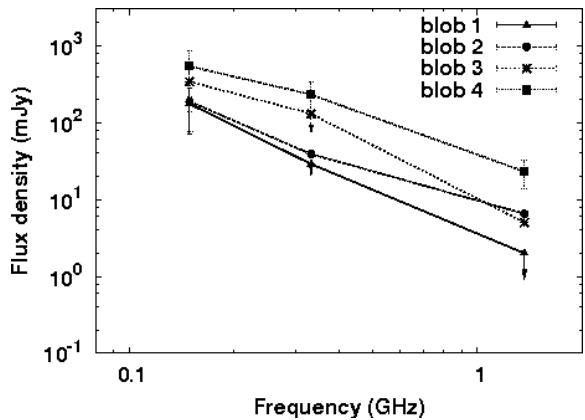}
\caption{Integrated spectra of blobs 1, 2, 3 and 4. For blob 1 upperlimits at 330 and 1363 MHz are plotted; for blobs 2 and 3 upperlimits at 330 MHz are plotted.}{\label{fig 3}}
\end{figure}
\begin{figure}
\epsscale{1}
\plotone{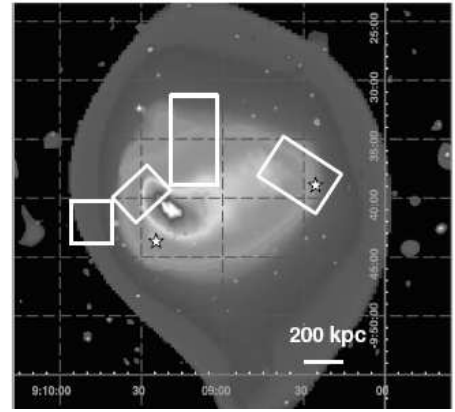}
\caption{A754 : X-ray image in the band 0.8-2 keV with XMM-Newton (Henry et al 2004). Positions of optical clumps of  galaxies (ZZ95) are marked by star symbols. The locations of the diffuse radio emissions blobs 1, 2, 3 and 4 are marked by rectangles.}{\label{fig 4}}
\end{figure}
\begin{figure}
\epsscale{1}
\plotone{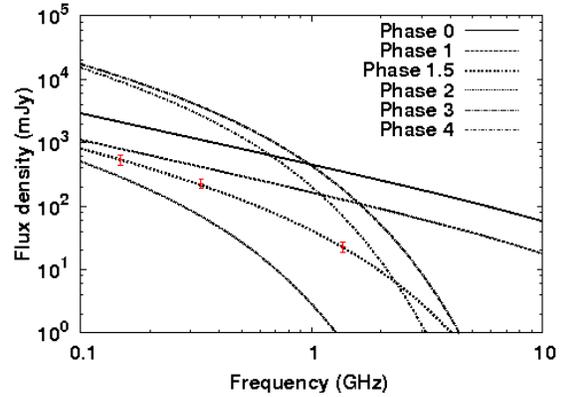}
\caption{Model spectra obtained using the adiabatic compression model (See Table 3 for the parameters). The model spectrum in phase 1.5 fits the observed spectrum (dots) of blob 4 best.}{\label{fig 5}}
\end{figure}
\clearpage
\begin{table}
\caption{Observations \label{tbl-1}}
\begin{tabular}{lccccc}
\tableline\tableline
Frequency  &Telescope  &Date&Time on&Synthesized  &rms \\
 (MHz)& &of obsn.&source(min)&beam&(mJybeam$^{-1}$)\\ 
 \hline
74(KA01)&VLA-C&21 Mar. 2000&180&$316''\times233''$&200\\
 150& GMRT &17 Sep. 2007 &225 &$21'' \times 21''$ &2.6 \\ 
 330& GMRT & 23 Jun. 2005&150&$10'' \times 10'' $ & 0.6   \\
330(KA01)& VLA-C & 21 Mar. 2000&180&$75''\times56''$ & 5.3  \\
1363(BA03) &VLA-D &25 Sep. 2000 & 180 &$50'' \times 50''$& 0.1 \\ 
 \tableline
\end{tabular}
\end{table}
\begin{table}
\caption{Flux densities of diffuse radio emission \label{tbl-2}}
\begin{tabular}{lcccccllll}
\tableline\tableline
  &\multicolumn{3}{c}{Integrated Flux density(mJy)}  &$\alpha^{150}_{1363}$& Extent  \\
  & (150 MHz)  &(330 MHz)&(1363 MHz)& &(kpc$^{2}$)\\ \hline
Blob 1         &174 &$<$28   & $<$2.0 &2.0&   240$\times$ 200    \\ 
Blob 2         &188 &$<$38   & 6.5 &1.5 &    260$\times$ 160  \\ 
Blob 3         &340 &$<$130   & 5.0 &1.9 & 280$\times$ 540      \\
Blob 4         &540 &230   &23.0            &1.4&350$\times$ 400 \\
\tableline
\end{tabular}
\end{table}
\begin{table}
\caption{Adiabatic compression model fit parameters\label{tbl-3}}
\begin{tabular}{llllll}
\tableline\tableline
          &$\Delta t$  &$\tau $ &$b$  & $V$       & $B$ \\
          &(Gyr)         &(Gyr)     &     &(Mpc$^{3}$)& ($\mu$G)\\ \hline
Phase 0   &0           &0.015   &1.8     & 0.02        & 1.0 \\ 
Phase 1   &0.0054      &0.01    &1.2     & 0.04        &  0.7    \\ 
Phase 1.5 &0.09        &$\infty$  & 0    & 0.04        & 0.7  \\ 
Phase 2   &0.1         &$\infty$  & 0    & 0.04        & 0.7 \\
Phase 3   &0.069       &-0.11   & 2.0    & 0.006        & 2.42 \\
Phase 4   &0.02        &$\infty$     &0     &0.006         & 2.42\\
\tableline
\end{tabular}
\end{table}
\end{document}